\documentclass[sigconf,natbib=true]{acmart}
\usepackage{xcolor}
\AtBeginDocument{%
  }

\copyrightyear{2026}
\acmYear{2026}
\setcopyright{acmlicensed}
\acmConference[Sensemaking @ CHI '26]{Sensemaking Workshop at CHI 2026}{April 13--17, 2026}{Barcelona, Spain}
\acmBooktitle{Sensemaking Workshop at CHI 2026, April 13--17, 2026, Barcelona, Spain}
\acmPrice{}
\acmISBN{}
\acmDOI{}

\begin{document}
\title{Envisioning Sensemaking in Multi-Human, Multi-Agent Collaborative Knowledge Work}

\author{Zhitong Guan}
\email{klarazt@utexas.edu}
\orcid{0009-0008-6950-2638}
\affiliation{%
  \institution{The University of Texas at Austin}
  %%\streetaddress{P.O. Box 1212}
  \city{Austin}
  \state{Texas}
  \country{USA}
  %%\postcode{43017-6221}
}

\author{Soo Young Rieh}
\email{rieh@ischool.utexas.edu}
\orcid{}
\affiliation{%
  \institution{The University of Texas at Austin}
  %%\streetaddress{P.O. Box 1212}
  \city{Austin}
  \state{Texas}
  \country{USA}
  %%\postcode{43017-6221}
}

\renewcommand{\shortauthors}{Guan and Rieh}

\begin{abstract}
    Sensemaking is central to knowledge work, where people search, evaluate, interpret, and use information over time to construct durable understanding. The rise of generative AI has begun to reshape this process: GenAI systems now perform interpretive functions such as summarization, synthesis, and thematic grouping that knowledge workers have traditionally carried out themselves. In collaborative settings, these shifts compound, complicating how teams divide interpretive labor, trust one another's contributions, and negotiate shared understanding. In this position paper, we examine how GenAI reshapes sensemaking in collaborative knowledge work and propose five design principles for multi-human, multi-agent collaborative sensemaking: dynamic multi-layer information representations, active identification and bridging of gaps in understanding, critical engagement with information, verifiability, and accountability. Building on these principles, we introduce a conceptual framework for a dynamic shared representational workspace in which knowledge workers and specialized AI agents jointly gather evidence, schematize, hypothesize, and pursue collaborative goals. Through a partner agent, a shared space agent, and an orchestrator agent, the framework preserves the provenance and authorship of contributions and traces the evolution of both individual and shared interpretations, supporting coherent, negotiated knowledge construction that current generative AI systems tend to obscure.
\end{abstract}


% %% the work being presented. Separate the keywords with commas.
% \keywords{}

\maketitle

\noindent\textbf{Author's Accepted Manuscript (AAM).} This is the Author's Accepted Manuscript version of the article: Guan, Z., \& Rieh, S. Y. (2026). Envisioning Sensemaking in Multi-Human, Multi-Agent Collaborative Knowledge Work. Accepted for publication in \textit{Sensemaking @ CHI 2026}.
\vspace{1em}

\section{Introduction}

Sensemaking is perhaps nowhere more consequential than in knowledge work. Across professional and scholarly domains, people engage in sustained sensemaking as they search, evaluate, interpret, and use information over time, moving between specific evidence and higher-level concepts, schemata, and hypotheses. Such workflows have traditionally supported continuity in reasoning and accountability in knowledge construction. The introduction of generative AI has transformed knowledge work workflows, placing new pressures on established sensemaking frameworks. As Generative AI (GenAI) systems increasingly perform interpretive functions such as generating summaries, syntheses, and thematic groupings within search systems, they assume roles historically carried out by knowledge workers. In collaborative settings, sensemaking becomes even more complex as team dynamics intersect with human–AI interaction.

In this position paper, we examine how AI reshapes sensemaking in collaborative knowledge work and propose five design principles for multi-human–AI collaborative sensemaking. Building on these principles, we introduce a conceptual framework for a dynamic representational workspace that enables individual sensemaking to contribute to collective sensemaking with accountability. Through specialized agents, the framework preserves the provenance and authorship of contributions and traces the evolution of both individual and shared interpretations, supporting coherent, negotiated knowledge construction that current generative AI systems tend to obscure.

\section{Challenges of Sensemaking with Generative AI in Collaborative Knowledge Work}

Sensemaking can be broadly understood as individual or collective construction of knowledge. Sensemaking has been examined across multiple disciplines, each adopting distinct perspectives and approaches. In HCI, \citet{russellCostStructureSensemaking1993} define it as searching for a representation and encoding data in that representation to answer task-specific questions through iterative cycles. \citet{pirolliSensemakingProcessLeverage2005} develop a descriptive model with a foraging loop and a sensemaking loop in which users cycle between gathering and filtering material and refining conceptual models that fit the evidence. In information science, \citet{dervinOverviewSensemakingResearch1983} frames sensemaking as a gap-bridging activity in which people construct meaning through information seeking and use, driven by evolving information needs. In cognitive systems engineering, \citet{kleinMakingSenseSensemaking2006} describe a bidirectional process in which frames determine what counts as relevant data while data reshape frames. In organizational communication, \citet{weickSensemakingOrganizations1995} situates sensemaking as inherently social and organizational, constructed collectively through interaction, narrative, and action rather than residing in any individual mind.

These disciplinary perspectives reveal differing emphases in how sensemaking is conceptualized and studied. For instance, Weick emphasizes retrospective sensemaking where people act first and construct meaning afterward, while Dervin positions it as a deliberate, prospective response to gaps in understanding. Other conceptual frameworks mainly regard sensemaking iterative. Rather than resolving these different emphases, our paper draws on their shared characterization of sensemaking as a constructive and dynamic process through which people develop workable understandings of complex information spaces and incoming events in order to act \cite{dervinOverviewSensemakingResearch1983, russellCostStructureSensemaking1993, pirolliSensemakingProcessLeverage2005, kleinMakingSenseSensemaking2006, weickSensemakingOrganizations1995}.

When sensemaking is distributed across multiple actors, additional complexities emerge. Collective sensemaking theories characterize groups as distributed cognitive systems in which meaning is constructed not within individual minds but across actors, tools, and shared representations \cite{weickSensemakingOrganizations1995, lubckeMultimodalCollectiveSensemaking2025}. Collective sensemaking proceeds through iterative cycles of individual interpretation, communication, and negotiation across multiple representational levels, producing provisional shared understanding that enables coordinated action \cite{cristofaroOrganizationalSensemakingSystematic2022}. \citet{sharmaArtifactUsefulnessUsage2009a} show that shared artifacts alone can support successful handoffs among team members, but that quality and timing matter. Early, unstructured materials are harder to build on than later, more coherent ones. These dynamics introduce further demands: managing dynamic and shared representations \cite{oddenDefiningSensemakingBringing2019}; negotiating and gatekeeping across actors \cite{lubckeMultimodalCollectiveSensemaking2025}; and accountability to evidence, to one another’s inputs, and to the shared norms of the community of practice \cite{sandbergSensemakingReconsideredBroader2020, brownMakingSenseSensemaking2015}.

\begin{figure*}[h]
  \centering
  \includegraphics[height=13.3cm]{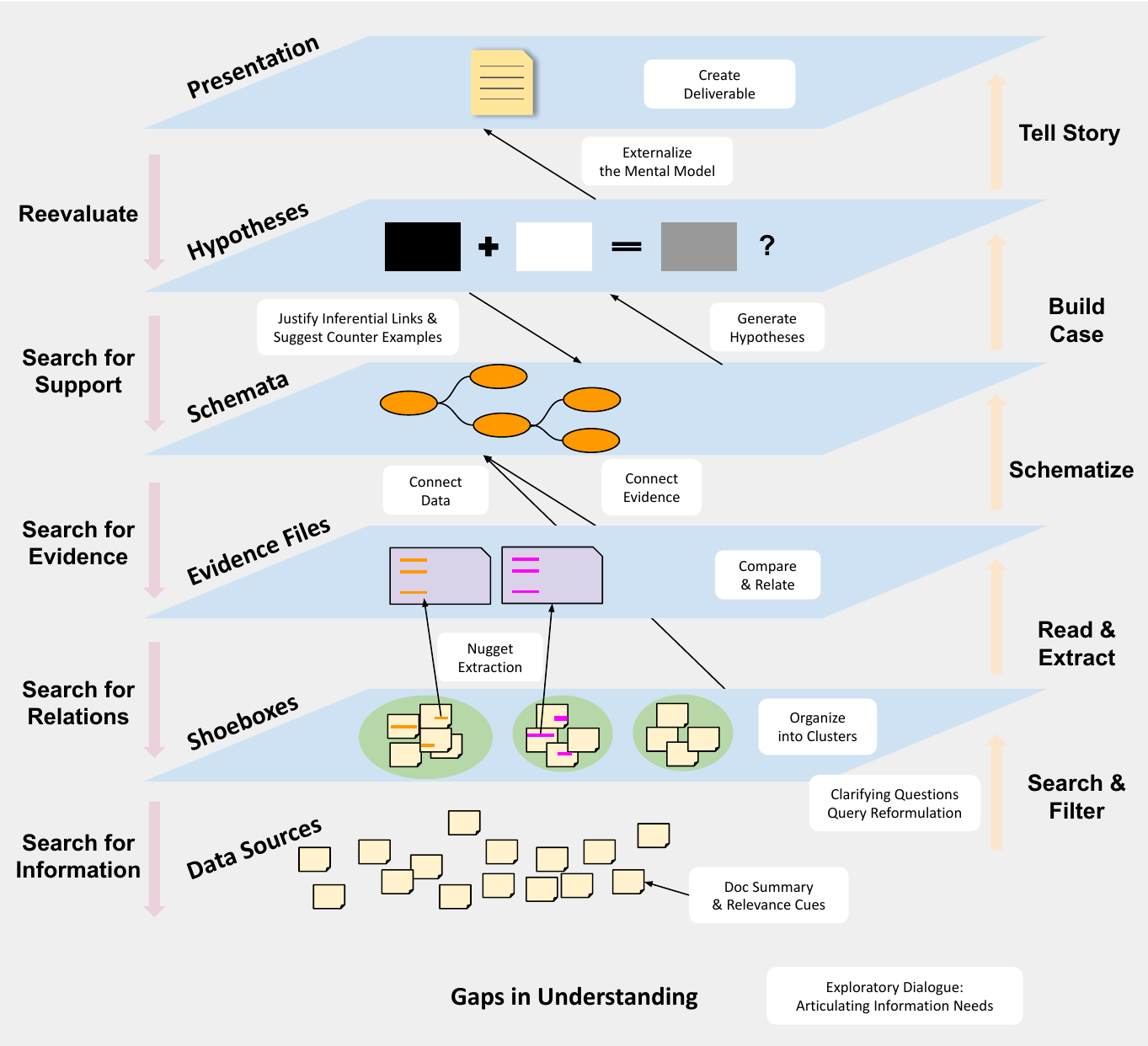}
  \caption{A shared multi-layered representational workspace for sensemaking in knowledge work, extending Pirolli and Card's model of sensemaking loop for analysis \cite{pirolliSensemakingProcessLeverage2005}. White annotation boxes identify sub-activities operating within and between representational layers which can be AI-mediated.}
  \Description{The figure illustrates a multi-layered representational workspace structured across six progressive layers: Data Sources, Shoeboxes, Evidence Files, Schemata, Hypotheses, and Presentation. Each layer represents a qualitatively different stage of information organization, from raw unstructured documents to refined communicable narratives. White annotation boxes with black arrows identify specific sub-activities within and between layers, such as nugget extraction, schematizing, hypothesis generation, and externalizing the mental model, reflecting the granular cognitive processes knowledge workers perform as they move through the workspace. The entire process is grounded in gaps in understanding at the base, which continuously motivate new information needs and re-entry into the foraging loop.}
  \label{fig:sensemakigloop}
\end{figure*}

These established sensemaking frameworks now face new pressures as GenAI systems take on interpretive roles traditionally performed by human actors. GenAI-powered systems generate summaries, syntheses, and thematic groupings directly within search and information use. When AI contributes interpretation rather than simply retrieval, the boundaries between human and machine sensemaking blur. Users may lose visibility into how data sources were selected, how evidence was weighted, how conflicting interpretations were handled, or what was excluded. For example, a traditional literature review requires evaluating search results, grouping papers into themes, comparing definitions and methods, and iteratively revising one’s conceptual structure as new evidence is encountered. A ready-made GenAI synthesis compresses this work into assessing output plausibility rather than building an interpretable, defensible understanding from the data sources. AI-mediated sensemaking can introduce hallucinations, undermine accountability, and raise concerns about whether novel insights can still emerge when interpretation is outsourced \cite{silvolaAImediatedSensemakingHigher2025}.

In collaborative contexts, these challenges compound further. Teams engaged in knowledge work routinely rely on division of labor and expertise, trusting one another's judgments based on reputation, domain knowledge, and established working relationships. But when people work not only with human collaborators, but also with AI systems performing interpretive labor, the basis for trust can become ambiguous: was AI used as a tool for analysis, or as a substitute for it? AI-mediated contributions can obscure both the interpretive process and the degree of human oversight \cite{coeckelberghNarrativeResponsibilityArtificial2023}. This new workflow risks undermining collaborative knowledge construction. Teams may spend more time re-verifying work they would normally trust, or worse, accept contributions without the critical negotiation that collective sensemaking requires. AI systems must therefore be designed and evaluated not just for output quality, but for how they sustain accountable, ongoing sensemaking across individuals and teams.

\begin{figure*}[h]
  \centering
  \includegraphics[width=\textwidth]{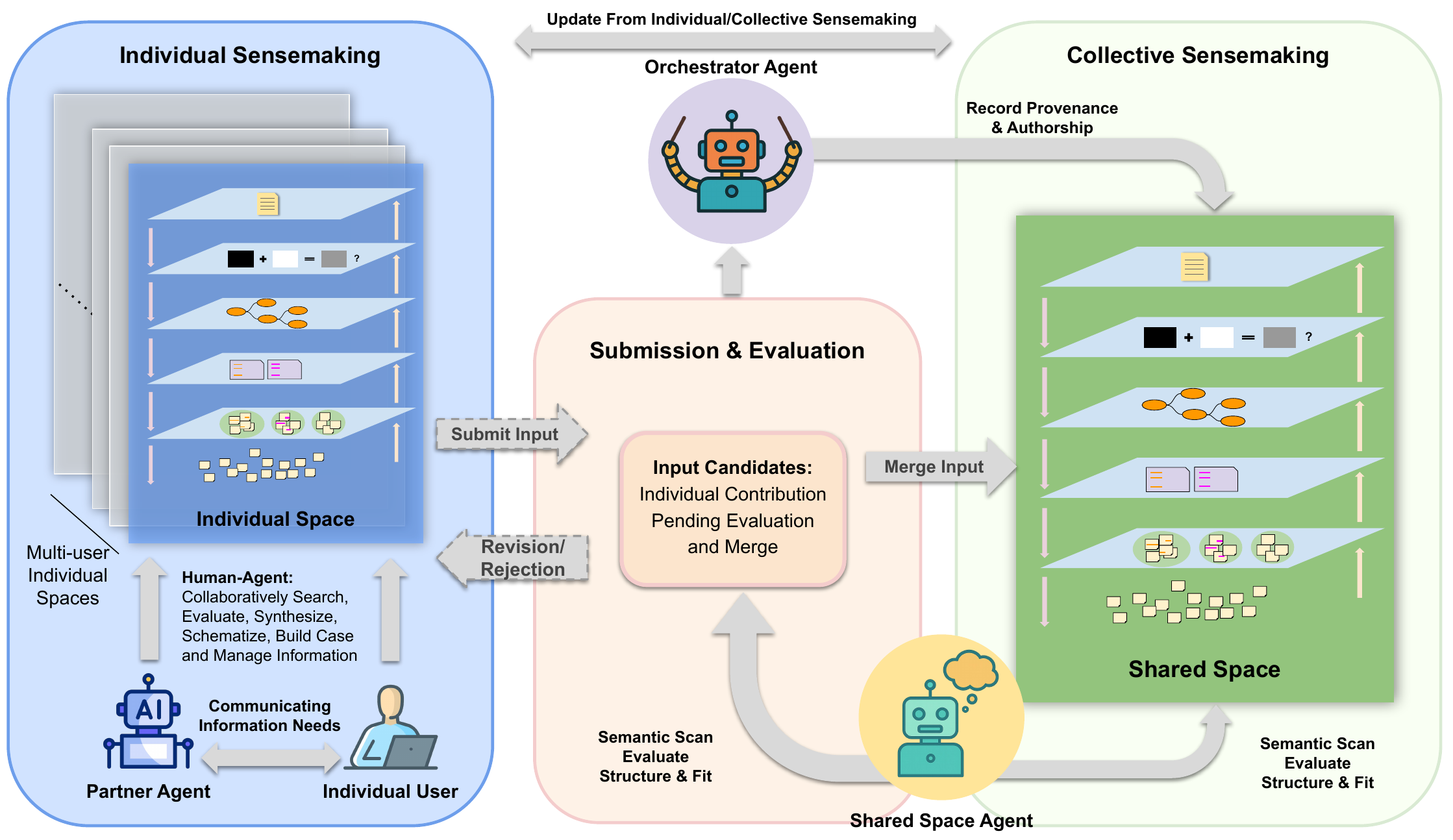}
  \caption{A Dynamic Conceptual Framework of a Shared Representational Workspace for Multi-Human, Multi-Agent Collaboration and Sensemaking. The framework comprises three zones: Individual Sensemaking (left, blue) where users work in individual spaces alongside a Partner Agent to explore and refine sensemaking; a Submission and Evaluation process (center, pink) where the Orchestrator Agent documents provenance and traces authorship; and Collective Sensemaking (right, green) where the Shared Space Agent semantically evaluates, integrates, and synchronizes contributions to the shared space.}
  \Description{Each user works within their Individual Space alongside a Partner Agent, which helps articulate information needs, retrieve relevant content, and prepare contributions for submission. Users retain autonomy to inspect, reorganize, and refine their representations independently. When ready, users submit contributions to the central Submission and Evaluation zone, where the orchestrator agent documents authorship and provenance, keeping the history of collaboration traceable. The Shared Space Agent then semantically evaluates the submission, determining where it fits within existing structures and what conceptual connections it introduces, before proposing how it should be integrated. Accepted contributions are merged into the Shared Space, which mirrors the same multi-layered representational structure as individual spaces. Updates propagate back to individual spaces through synchronization and integrity checks, supporting alignment without imposing uniformity. When contributions conflict or overlap ambiguously, the Shared Space Agent preserves them as distinct proposals, keeping alternative interpretations visible and negotiable rather than resolving them automatically.}
  \label{fig:framework}
\end{figure*}

\section{Design Principles for Multi Human-AI Collborative Sensemaking Support}

We argue that these challenges call for rethinking the structure of the workspace in which individual and collective sensemaking now occurs. As GenAI becomes embedded in knowledge work, sensemaking is no longer enacted only at the level of human–information interaction, but increasingly through intertwined human–AI, AI–AI, and human–human interactions. This shift demands systems that support the continuous interplay between individual interpretation and collective knowledge construction, while preserving visibility into how understanding develops across actors and over time. Prior work established that shared representations, negotiated meaning, and accountability are essential to collective understanding \cite{weickSensemakingOrganizations1995, lubckeMultimodalCollectiveSensemaking2025}. Yet existing systems treat AI as a peripheral assistant rather than an embedded participant in sensemaking. We propose five design principles for tools that support sensemaking in multi-human, multi-agent collaborative knowledge work as follows.

\textbf{D1 Dynamic multi-layer information representations}: help externalizing understanding into structured, editable representations spanning multiple levels of abstraction.
    
\textbf{D2 Active identification and bridging of gaps in understanding}: help recognize what is not yet understood, and the construction of meaning toward it, remains visible and supported rather than bypassed.
    
\textbf{D3 Critical engagement with information}: help evaluate, reason, and actively reorganize data sources and interpretations rather than being limited to accepting ready-made conclusions.
    
\textbf{D4 Verifiability}: help trace, justify, and explain the interpretations they develop: \textit{can I explain how I arrived at this understanding?}
    
\textbf{D5 Accountability}: help make inputs visible and comparable across collaborators. Interpretations remain open to dialog and revision, and the evolution of shared understanding can be traced over time.

\section{A Dynamic Shared Representational Workspace for Collaboration and Sensemaking}

These principles point to an opportunity for designing systems that support individual and collective sensemaking for collaborative knowledge work within a workspace shared by humans and AI agents. We propose a conceptual framework for a dynamic shared representational workspace in which knowledge workers and AI agents jointly gather evidence, schematize, hypothesize, and pursue collaborative task goals, as depicted in Figure \ref{fig:framework}. To make the framework concrete, we illustrate it through a running scenario of a financial market analysis task.

As shown in Figure \ref{fig:sensemakigloop}, the multi-layer representation enables people to organize information spanning raw data, excerpts, evidence files, schemata, hypotheses, and presentations (\textbf{D1}), reflecting how people move between detail and abstraction during sensemaking. 

\textit{Scenario. A financial analyst starts with a need to better understand a possible shift in the volatility regime. She gathers data sources including trading volume, prices, macro indicators, and news sentiment, clustering them into shoeboxes by year. A recurring co-movement between widening credit spreads and equity drawdowns in the sector begins to stand out. She isolates specific historical episodes where this pattern holds as evidence and uses them to form a working hypothesis: when credit spreads widen beyond a threshold, a short position in the sector’s equity index is likely to be profitable.}

The individual sensemaking is also supported through human--AI collaboration. Each individual workspace includes a \textbf{partner agent} that helps knowledge workers navigate the information foraging and sensemaking loops (Figure \ref{fig:framework}). Directly supporting the identification and bridging of gaps in understanding (\textbf{D2}), \textbf{the partner agent} helps them articulate and refine information needs, retrieve relevant content, reframe ambiguous ideas, and generate alternative formulations. It can also support comparison by surfacing alternative interpretations and identifying ambiguities in reasoning (\textbf{D4}). The human actor remains the primary reasoner, actively evaluating agent-generated syntheses, reorganizing evidence, and revising interpretations rather than treating them as conclusions (\textbf{D3}).

\textit{Scenario. As the analyst clusters signals into shoeboxes, her partner agent notices she has not yet considered options market data as a potential source and surfaces it as a relevant addition. When she forms her credit spread hypothesis, the agent flags a gap in her reasoning for not accounting for episodes where spreads widened without a corresponding equity drawdown, prompting her to refine her condition for the hypothesis.}

When an individual knowledge worker is satisfied with their sensemaking, they can submit inputs into the shared space to update their collaborators on what they have learned. These inputs may range from new raw data sources to a more developed chain of sensemaking, including evidence and hypotheses. Rather than incorporating them into the shared space automatically, \textbf{the shared space agent} treats them as candidate additions (\textbf{D5}). It conducts a semantic scan of the input, evaluates it in relation to the existing collective sensemaking, and determines where it fits and what new representational connections it may introduce before proposing how it should be integrated. In parallel, \textbf{the orchestrator agent} ensures accountability (\textbf{D5}) by recording the authorship and provenance of each input so that the history of individual contributions and collaboration remains visible and traceable (Figure \ref{fig:framework}). This division of labor ensures that collective sensemaking is constructed through negotiation, and that proposed additions can be rejected, revised, or reversed.

\textit{Scenario. The researcher submits her credit spread hypothesis into the shared workspace. The shared space agent connects it to a related schema a colleague has been developing around macro regime shifts, while the orchestrator agent records her authorship and the evidence chain behind her submission.}

When submitted inputs introduce competing interpretations or challenge an emerging hypothesis, \textbf{the shared space agent} preserves them as separate candidates rather than resolving them immediately. By keeping alternatives visible and linking them to their supporting evidence, the workspace helps collaborators trace and evaluate interpretations before making decisions (\textbf{D4}). Collaborators can examine why an input was proposed, what evidence supports it, and how it relates to existing structures before deciding whether to revise, branch, defer, or integrate it. At the individual level, \textbf{the partner agent} can further help inspect the chain of reasoning and clarify how it connects to prior evidence before the input is accepted or revised, further supporting verifiability (\textbf{D4}).

\textit{Scenario. A colleague submits a revised hypothesis assigning greater predictive weight to earnings revisions. The shared space agent preserves both as separate candidates. The analyst uses her partner agent to inspect how her colleague's evidence chain.}

Because individual and shared views draw on the same underlying representational structure, dynamic interfaces keep them synchronized as the collaborative workspace evolves (\textbf{D1}). Each individual space contains the same multi-layered representational structure as the shared space, while allowing users to work through personalized views and at different levels of abstraction. When a contribution is committed, the shared space updates the relevant layers while maintaining coherence with existing structures, which can also be propagated back to other individual spaces. Even when collaborators work at different representational layers and in individual spaces, the shared workspace provides a coherent backbone that keeps their interpretations legible in relation to one another. As a result, collaborators can see how others' insights connect to their own, trace the evolution of ideas, and handoff sensemaking without the fragmentation typical of today's multi-tool workflows.

%% the bibliography file
% ACM recommended settings to shorten author lists
\bibliographystyle{ACM-Reference-Format}
\bibliography{references}

% \appendix

\end{document}